\documentclass[12pt]{iopart}
\usepackage{graphicx}
\usepackage{url}

\begin{document}

\topical{Ho\v{r}ava-Lifshitz Cosmology: A Review}
\author{Shinji Mukohyama}
\address{
Institute for the Physics and Mathematics of the Universe (IPMU)\\ 
The University of Tokyo\\
5-1-5 Kashiwanoha, Kashiwa, Chiba 277-8582, Japan
}

\begin{abstract}
 This article reviews basic construction and cosmological implications
 of a power-counting renormalizable theory of gravitation recently
 proposed by Ho\v{r}ava. We explain that (i) at low energy this theory
 does not exactly recover general relativity but instead mimic general
 relativity plus dark matter; that (ii) higher spatial curvature terms
 allow bouncing and cyclic universes as regular solutions; and that
 (iii) the anisotropic scaling with the dynamical critical exponent
 $z=3$ solves the horizon problem and leads to scale-invariant
 cosmological perturbations even without inflation. We also comment on
 issues related to an extra scalar degree of freedom called scalar
 graviton. In particular, for spherically-symmetric, static, vacuum
 configurations we prove non-perturbative continuity of the 
 $\lambda\to 1+0$ limit, where $\lambda$ is a parameter in the kinetic
 action and general relativity has the value $\lambda=1$. We also derive
 the condition under which linear instability of the scalar graviton
 does not show up.
 \begin{flushright}
  (IPMU10-0120)
 \end{flushright}
\end{abstract}

\tableofcontents

\pagestyle{plain}

\section{Introduction}

One of the biggest difficulties in attempts toward the theory of quantum
gravity is the fact that general relativity is non-renormalizable. This
would imply loss of theoretical control and predictability at high
energies. In January 2009, Ho\v{r}ava proposed a new theory of gravity
to evade this difficulty by invoking a Lifshitz-type anisotropic scaling
at high energy~\cite{Horava:2009uw}. This theory, often called
Ho\v{r}ava-Lifshitz gravity, is power-counting renormalizable and is
expected to be renormalizable and unitary. Having a new candidate theory
for quantum gravity, it is important to investigate its cosmological
implications.

There are a number of interesting cosmological implications of
Ho\v{r}ava-Lifshitz gravity. For example, higher spatial curvature terms 
lead to regular bounce solutions in the early
universe~\cite{Calcagni:2009ar,Brandenberger:2009yt}. Higher curvature 
terms might also make the flatness problem
milder~\cite{Kiritsis:2009sh}. The anisotropic scaling with
$z=3$ solves the horizon problem and leads to scale-invariant
cosmological perturbations without
inflation~\cite{Mukohyama:2009gg}. The anisotropic scaling provides a
new mechanism for generation of primordial magnetic seed
field~\cite{Maeda:2009hy}, and also modifies the spectrum of
gravitational wave background via a peculiar scaling of radiation energy
density~\cite{Mukohyama:2009zs}. In parity-violating version of the 
theory, circularly polarized gravitational waves can also be generated
in the early universe~\cite{Takahashi:2009wc}. The lack of local
Hamiltonian constraint leads to dark matter as an integration 
``constant''~\cite{Mukohyama:2009mz,Mukohyama:2009tp}.

The purpose of this article is to review basic construction of the
theory and some of its cosmological implications. In
Sec.~\ref{sec:HLgravity} we explain basics of Ho\v{r}ava-Lifshitz 
gravity, such as power-counting argument, symmetry, basic quantities,
action and equations of motion. In  Sec.~\ref{sec:scalargraviton} we
comment on issues related to an extra scalar degree of freedom called
scalar graviton, and consider the limit in which general relativity is
supposed to be recovered. We explicitly see the known result that the 
naive metric perturbation breaks down in this limit for the scalar
graviton. However, this does not necessarily imply the loss of
predictability. Indeed, for spherically-symmetric, static, vacuum
configurations we shall show that the limit is non-perturbatively
continuous. This result might correspond to what is called Vainshtein 
mechanism~\cite{Vainshtein:1972sx} in theories of massive
gravity~\cite{Fierz:1939ix} and suggest that the extra scalar degree of
freedom might safely decouple from the rest of the world in the limit.
In Sec.~\ref{sec:cosmology} we shall review some of cosmological
implications: the dark matter as an integration ``constant'', bouncing
and cyclic universes and generation of scale-invariant cosmological
perturbation without inflation. Finally, Sec.~\ref{sec:summary} is
devoted to a summary of this article and some discussions.

\section{Ho\v{r}ava-Lifshitz gravity}
\label{sec:HLgravity}

\subsection{Preliminaries}

\subsubsection{Power-counting}

Let us begin with heuristically explaining the usual power-counting
argument in field theory. As the simplest example, let us consider a
scalar field with the canonical kinetic term: 
%
\begin{equation}
 \frac{1}{2}\int dt d^3\vec{x}\ \dot{\phi}^2, 
  \label{eqn:scalar-kineticterm}
\end{equation}
where an overdot represents a time derivative. The scaling dimension of
the scalar field $\phi$ is determined by demanding that the kinetic term
be invariant under the scaling
%
\begin{equation}
 t \to b t, \quad \vec{x} \to b x, \quad \phi\to b^{-s}\phi, 
  \label{eqn:scaling}
\end{equation}
where $b$ is an arbitrary number and $s$ is the scaling dimension to be
determined. The invariance of the kinetic term under the scaling leads
to the condition
%
\begin{equation}
 1+3-2-2s = 0,
\end{equation}
where $1$ comes from $dt$, $3$ from $d^3\vec{x}$, $-2$ from two time
derivatives and $-2s$ from two $\phi$'s. Thus, we obtain $s=1$. In other
words, the scalar field scales like energy. With this scaling in mind,
it is easy to see that an $n$-th order interaction term behaves as 
%
\begin{equation}
 \int dtd^3\vec{x} \phi^n \propto E^{-(1+3-ns)},
\end{equation}
where $E$ is the energy scale of the system of interest. Here, the minus
sign in the exponent comes from $-1$ in $E\to b^{-1}E$, $1$ in the
parentheses comes from $dt$, $3$ from $d^3\vec{x}$ and $-ns$ from
$\phi^n$. Now, it is expected that we have a good theoretical control of
ultraviolet (UV), i.e. high $E$, behaviors if the exponent is
non-positive. Since $s=1$, this condition leads to $n\leq 4$. This is
the power-counting renormalizability condition.

We are interested in gravity. Unfortunately, Einstein gravity is not
power-counting renormalizable. This is because the curvature is a highly
nonlinear functional of the metric and there are graviton interaction
terms with $n$ higher than $4$. The non-renormalizability is one of
difficulties in attempts to quantize general relativity.

\subsubsection{Abandoning Lorentz symmetry}

As already stated, Ho\v{r}ava-Lifshitz gravity is power-counting
renormalizable. How does it evade the above argument? The basic idea is
very simple but potentially dangerous: abandoning Lorentz symmetry and
invoking a different kind of scaling in the UV. The scaling invoked
here, often called {\it anisotropic scaling} or {\it Lifshitz scaling}, is 
%
\begin{equation}
 t \to b^z t, \quad \vec{x} \to b x, \quad \phi\to b^{-s}\phi, 
  \label{eqn:anisotropic-scaling}
\end{equation}
where $z$ is a number called {\it dynamical critical exponent}. 

Let us now see how the power-counting argument changes if the scaling is
anisotropic as in (\ref{eqn:anisotropic-scaling}). Invariance of the
canonical kinetic term (\ref{eqn:scalar-kineticterm}) under this scaling
leads to 
%
\begin{equation}
 z + 3 - 2z - 2s = 0,
\end{equation}
where $z$ comes from $dt$, $3$ from $d^3\vec{x}$, $-2z$ from two time
derivatives and $-2s$ from two $\phi$'s. Then we obtain
%
\begin{equation}
 s = \frac{3-z}{2}.
  \label{eqn:scaling-dim}
\end{equation}
This of course recovers the previous result $s=1$ for $z=1$. What is 
interesting here is that $s=0$ if $z=3$. This implies that, if $z=3$,
the amplitude of quantum fluctuations of $\phi$ does not change as the
energy scale of the system changes. The $n$-th order interaction term
behaves as
%
\begin{equation}
 \int dtd^3\vec{x} \phi^n \propto E^{-(z+3-ns)/z},
\end{equation}
where $-1/z$ in the exponent comes from $-z$ in $E\to b^{-z}E$, $z$ in
the parentheses comes from $dt$, $3$ from $d^3\vec{x}$ and $-ns$ from
$\phi^n$. For $z=3$ (and thus $s=0$), the exponent is negative for any 
$n$ and, therefore, any nonlinear interactions are power-counting
renormalizable. For $z>3$, the theory is power-counting
super-renormalizable.

From the above consideration, it is expected that gravity may become 
renormalizable if the anisotropic scaling with $z\geq 3$ is realized in
the UV.

\subsubsection{Scalar field action}
\label{subsubsec:scalaraction}

We would like to realize the anisotropic scaling with $z\geq 3$ in the
UV to construct renormalizable nonlinear theories. On the other
hand, in order to recover the Lorentz invariance in the infrared (IR),
we would like to realize the usual scaling with $z=1$ at low energy. A
simple example is a scalar field with the following free-part action: 
%
\begin{equation}
 I_{free} = \frac{1}{2}\int dtd^3\vec{x}
  \left(\dot{\phi}^2+\phi {\cal O}\phi\right),
  \label{eqn:free-scalar-flat}
\end{equation}
where 
%
\begin{equation}
 {\cal O} = \frac{\Delta^3}{M^4}
  - \frac{\kappa\Delta^2}{M^2} + c_{\phi}^2\Delta - m_{\phi}^2, 
\end{equation}
$M$ is the energy scale corresponding to the transition from the
$z=1$ scaling to the $z=3$ scaling, $\kappa$ is a constant, $c_{\phi}$
is the sound speed, i.e. the limit of speed in the IR, $m_{\phi}$
is the mass of the field, and $\Delta$ is the Laplacian in the
$3$-dimensional space~\footnote{If photons have this kind of dispersion
relation with $\kappa=O(1)$, $c_{\phi}=1$ and $m_{\phi}=0$ then
experiments such as Fermi GBM/LAT~\cite{:2009zq} and
MAGIC~\cite{Albert:2007qk} set a lower bound on $M$ as
$M>10^{11}GeV$.}.

In the UV, the sixth-order spatial derivative term dominates over
lower-order terms and balances with the time kinetic term which includes
two time derivatives. This naturally leads to the $z=3$ scaling. On the 
other hand, in the IR, the second-order spatial derivative term
and the mass term are dominant and, thus, the $z=1$ scaling is
realized. In this way, it is possible to realize the $z=3$ scaling in
the UV and the $z=1$ scaling in the IR.

However, one must be aware that all ``constants'' in the action are
subject to running under the renormalization group (RG) flow. Of course,
the sound speed is not an exception. If we consider many fields then the
sound speed for each field should run under the RG
flow~\cite{Iengo:2009ix}. We need a mechanism or symmetry to make sound
speeds of different species to be essentially the same at low
energies. More generally speaking, we need a mechanism or symmetry to
suppress Lorentz violating operators at low energies. Perhaps, embedding
the theory into a larger theory is necessary. One such possibility is
related to supersymmetry~\cite{GrootNibbelink:2004za}.

\subsection{Symmetry}

As explained in the previous subsection, the way the power-counting
renormalizability is achieved is to violate the Lorentz invariance and
to invoke the anisotropic scaling with the dynamical critical exponent
$z\geq 3$. Since the Lorentz invariance is not respected, we treat the
time coordinate $t$ and the spatial coordinates $\vec{x}^i$ ($i=1,2,3$)
separately.

The fundamental symmetry of the theory is the invariance under
space-independent time reparametrization and time-dependent spatial
diffeomorphism: 
%
\begin{equation}
 t \to t'(t), \quad \vec{x} \to \vec{x}'(t,\vec{x}).
  \label{eqn:foliation-preserving-diffeo}
\end{equation}
The time-dependent spatial diffeomorphism allows an arbitrary change of
spatial coordinates on each constant time surface. However, the time
reparametrization here is not allowed to depend on spatial
coordinates. As a result, unlike general relativity, in
Ho\v{r}ava-Lifshitz gravity the foliation of spacetime by constant time
hypersurfaces is not just a choice of coordinates but is a physical
entity. Indeed, the foliation is preserved by the symmetry transformation
(\ref{eqn:foliation-preserving-diffeo}). For this reason, the map
(\ref{eqn:foliation-preserving-diffeo}) is called {\it foliation
preserving diffeomorphism}.

In addition to the foliation preserving diffeomorphism invariance, we
assume that the theory is invariant under the {\it spatial parity}
$\vec{x}\to -\vec{x}$~\cite{Sotiriou:2009gy} and the 
{\it time reflection} $t\to -t$.

Finally, in order to render the theory power-counting renormalizable, we
would like to realize the anisotropic scaling with $z\geq 3$ at high
energy. In the present article, for concreteness, we mainly focus on the
case with $z=3$.

\subsection{Basic quantities and projectability condition}

Basic quantities of Ho\v{r}ava-Lifshitz gravity are 
\begin{equation}
\mbox{lapse}: N(t), \quad
\mbox{shift}: N^i(t,\vec{x}), \quad
\mbox{$3$d metric}: g_{ij}(t,\vec{x}),
\label{eqn:basicquantities}
\end{equation}
from which we can construct a four-dimensional spacetime metric of the
ADM form as
\begin{equation}
 ds^2 = -N^2dt^2 + g_{ij}(dx^i+N^idt)(dx^j+N^jdt).
  \label{eqn:ADMmetric}
\end{equation} 
While the shift $N^i$ and the $3$d metric $g_{ij}$ depend on both the
time coordinate $t$ and the spatial coordinates $\vec{x}$, the lapse $N$
is assumed to be a function of the time only. This condition on the
lapse is called the {\it projectability condition}.

The projectability condition stems from the foliation preserving
diffeomorphism. The lapse represents a gauge freedom associated with the
space-independent time reparametrization $t\to t'(t)$ and, thus, it is
fairly natural to restrict it to be
space-independent~\footnote{Abandoning the projectability condition
leads to phenomenological obstacles~\cite{Charmousis:2009tc} and 
theoretical inconsistency~\cite{Henneaux:2009zb}. On the other hand,
the criticisms made in \cite{Charmousis:2009tc,Henneaux:2009zb} do not 
apply if the projectability condition is respected.}. Of course, the
projectability condition is compatible with the foliation preserving 
diffeomorphism. The transformation of the basic quantities
(\ref{eqn:basicquantities}) under the infinitesimal foliation preserving
diffeomorphism, 
\begin{equation}
 \delta t = f(t), \quad \delta \vec{x}^i = \xi^i(t,\vec{x}),
\end{equation}
is defined as follows. 
\begin{eqnarray}
 \delta N & = & \partial_t(N f),   \nonumber\\
 \delta N^i & = & \partial_t (N^i f) + \partial_t \xi^i
  + {\cal L}_{\xi}N^i, \nonumber\\
 \delta (N_i) & = & \partial_t (N_i f) + g_{ij}\partial_t \xi^j
  + {\cal L}_{\xi}N_i, \nonumber\\
 \delta g_{ij} & = & f\partial_t g_{ij} + {\cal L}_{\xi}g_{ij},
  \label{eqn:infinitesimal-tr}
\end{eqnarray}
where $N_i=g_{ij}N^j$. Note that $\delta N$ is independent of spatial
coordinates since $f$ and $N$ are functions of time only. Thus the
projectability condition is compatible with the foliation preserving
diffeomorphism: the foliation preserving diffeomorphism maps a
space-independent $N$ to a space-independent $N$.

The equation of motion for the lapse corresponds to the generator of the
time reparametrization and is called the {\it Hamiltonian
constraint}. Since the lapse is independent of spatial coordinates, its
variations are also space-independent. This means that the Hamiltonian
constraint in Ho\v{r}ava-Lifshitz gravity is not a local equation but an
equation integrated over a whole space. In subsection
\ref{subsec:darkmatter} we shall discuss cosmological implication of the
global nature of the Hamiltonian constraint.

\subsection{Action}
\label{subsec:action}

The theory should respect the foliation preserving diffeomorphism. We
can then use the following ingredients in the action:
\begin{equation}
 Ndt, \quad \sqrt{g}d^3\vec{x}, \quad g_{ij}, 
  \quad D_i, \quad R_{ij},
\end{equation}
where $g$ is the determinant of $g_{ij}$, $D_i$ is the $3$-dimensional
covariant derivative compatible with $g_{ij}$ and $R_{ij}$ is the Ricci
tensor of $g_{ij}$. Note that the Ricci tensor includes all information
about the Riemann tensor since Weyl tensor identically vanishes in
$3$-dimensions.

\subsubsection{The UV action}

We should include time-derivative of the $3$-dimensional metric in the
action in order to make the metric dynamical. However, $\dot{g}_{ij}$ is
not covariant under the spatial diffeomorphism and, therefore,
$\dot{g}_{ij}$ should appear in the action as a part of the covariant
quantity called extrinsic curvature, 
\begin{equation}
 K_{ij} = \frac{1}{2N}\left(\dot{g}_{ij}-D_iN_j-D_jN_i\right).
\end{equation}
The extrinsic curvature transforms as a second-rank symmetric tensor
under the spatial diffeomorphism and as a scalar under the time
reparametrization. The time kinetic term for the metric is obtained by
squaring the extrinsic curvature and properly contracting indices. There
are two ways to contract indices: 
\begin{equation}
 I_{kin} = \frac{1}{16\pi G}\int Ndt \sqrt{g}d^3\vec{x}
  \left(K^{ij}K_{ij}-\lambda K^2 \right), \label{eqn:kinetic-action}
\end{equation}
where $G$ and $\lambda$ are constants, and $K=K^i_i$. In general 
relativity, $\lambda$ is fixed to $1$ because of higher symmetry. On the
other hand, in Ho\v{r}ava-Lifshitz gravity, any value of $\lambda$ is
compatible with the foliation preserving diffeomorphism invariance and
thus $\lambda$ is not fixed. We shall not include terms including
derivatives of the extrinsic curvature in the action. This is consistent
if the theory without those higher derivative terms is renormalizable:
those terms would be non-renormalizable and thus would not be generated
by quantum correction. For the same reason, we shall not include terms
cubic or higher order in the extrinsic curvature.

Invariant terms made of time derivatives of the shift would inevitably
include second or higher time derivatives of the spatial metric. For the
reason explained above, we shall not include those higher derivative
terms in the action. Time derivative of the lapse corresponds to the
connection in $1$-dimension spanned by the time but the curvature in
$1$-dimension is always zero. Thus, there is no invariant term made of
time derivatives of the lapse. Of course, the spatial derivative of the
lapse vanishes because of the projectability condition.

Since terms in the kinetic action (\ref{eqn:kinetic-action})
include two time derivatives, we should include terms with six spatial
derivatives in order to realize the $z=3$ scaling in the UV. (For a
general choice of $z$ ($\geq 3$) in the UV, we should include terms
with $2z$ spatial derivatives.) The foliation preserving diffeomorphism
invariance allows five such terms in the action,
\begin{eqnarray}
 I_{z=3} & = & \int Ndt \sqrt{g}d^3\vec{x}
  \left[ c_1 D_iR_{jk}D^iR^{jk}
   \right.
 \nonumber\\
 & & \left.
      + c_2 D_iRD^iR
      + c_3 R_i^jR_j^kR_k^i + c_4 R R_i^jR_j^i + c_5 R^3
     \right],
\end{eqnarray}
where $c_i$ ($i=1,\cdots,5$) are constants. Note that the other possible
term $D_iR_{jk}D^jR^{ki}$ is a linear combination of the above terms up
to total derivative and, thus, does not have to be included
explicitly. We do not include terms with more than six spatial
derivatives since they would be non-renormalizable and thus would not be
generated by quantum corrections if the theory without them is
renormalizable.

\subsubsection{Relevant deformations in the IR}

In the IR, terms with less number of spatial derivatives in the action
become important. There are two independent terms with four spatial
derivatives 
\begin{equation}
 I_{z=2} = \int Ndt \sqrt{g}d^3\vec{x}
  \left[ c_6 R_i^jR_j^i + c_7 R^2 \right],
\end{equation} 
one term with two spatial derivatives
\begin{equation}
 I_{z=1} = c_8 \int Ndt \sqrt{g}d^3\vec{x}\ R, 
\end{equation} 
and a constant
\begin{equation}
 I_{z=0} = c_9 \int Ndt \sqrt{g}d^3\vec{x}\ ,
\end{equation} 
where $c_i$ ($i=6,\cdots,9$) are constants.

We have written down all possible terms consistent with the symmetry of
the theory except for terms involving more than two time derivatives and
terms with more than six spatial derivatives. As already stated, those
higher-derivative terms excluded in the above construction would be
non-renormalizable and, thus, would not be generated by quantum
corrections if the theory without them is renormalizable. The theory
defined in this way is power-counting renormalizable and, thus, expected
to be renormalizable although renormalizability beyond the
power-counting argument has not been proved. Also, the theory is
expected to be unitary since the action does not include more than two
time derivatives. Note that the constants $G$, $\lambda$ and $c_i$
($i=1,\cdots,9$) are subject to running under the RG flow.

\subsubsection{IR action with $z=1$}

In the UV the theory naturally exhibits the $z=3$ scaling as the second
time derivative terms $I_{kin}$ and the sixth spatial derivative terms
$I_{z=3}$ balance with each other.

On the other hand, in the IR the forth and sixth spatial derivative
terms, $I_{z=2}$ and $I_{z=3}$, are unimportant. We therefore have the
following action describing the IR behavior of the theory: 
\begin{eqnarray}
 I_{IR} & = & I_{kin}+I_{z=1}+I_{z=0} \nonumber\\
 & = & \frac{M_{Pl}^2}{2}\int Ndt\sqrt{g}d^3\vec{x}
  \left( K^{ij}K_{ij}-\lambda K^2 + R-2\Lambda\right),
  \label{eqn:IRaction}
\end{eqnarray}
where $M_{Pl}^2\equiv 1/(8\pi G)$ and $\Lambda\equiv -8\pi Gc_9$, 
and we have set $16\pi G c_8$ to unity by rescaling of the time 
coordinate. This IR action naturally exhibits the $z=1$ 
scaling. Moreover, the action looks identical to the Einstein-Hilbert
action in the ADM formalism if $\lambda\to 1$. There are however two
important differences: (i) $\lambda$ does not have to be $1$ and is
subject to running under the RG flow; (ii)  the projectability condition
restricts the lapse $N$ to be a function of the time only. Regarding
(i), the RG flow of the theory has not been investigated and, thus, we
do not know whether $\lambda=1$ is an IR fixed point of the RG flow or
not. On the other hand, we shall discuss cosmological implication of the 
point (ii) in subsection \ref{subsec:darkmatter}.

\subsection{Equations of motion}

Adding the matter action $I_m$, the total action is
\begin{eqnarray}
 I & = &I_g+I_m, \\
 I_g & = & \frac{M_{Pl}^2}{2}
  \int Ndt \sqrt{g}d^3\vec{x}
  (K^{ij}K_{ij}-\lambda K^2+\Lambda+R+L_{z>1}),\\
 \frac{M_{Pl}^2}{2}
  L_{z>1} & = & (c_1 D_iR_{jk}D^iR^{jk}
  + c_2 D_iRD^iR + c_3 R_i^jR_j^kR_k^i
  \nonumber\\
  & &  + c_4 R R_i^jR_j^i + c_5 R^3) + (c_6 R_i^jR_j^i + c_7 R^2).
  \label{eqn:gravaction}
\end{eqnarray}
Here, we have rescaled the time coordinate to set $16\pi Gc_8$ to
unity. Note that not only the gravitational action $I_g$ but also the
matter action $I_m$ should be invariant under the foliation-preserving
diffeomorphism.

By variation of the action with respect to the lapse $N(t)$, we obtain
the Hamiltonian constraint 
\begin{equation}
 H_{g\perp}+H_{m\perp}=0, \label{eqn:HamiltonianConstraint}
\end{equation}
where
\begin{equation}
 H_{g\perp} \equiv -\frac{\delta I_g}{\delta N}
  = \int d^3\vec{x} {\cal H}_{g\perp}, \
 H_{m\perp} \equiv -\frac{\delta I_m}{\delta N},
\end{equation} 
and
\begin{equation}
  {\cal H}_{g\perp} = \frac{M_{Pl}^2}{2}\sqrt{g}
  (K^{ij}p_{ij}-\Lambda-R-L_{z>1}),  \quad
  p_{ij} \equiv K_{ij} - \lambda Kg_{ij}. 
\end{equation}
Variation with respect to the shift $N^i(t,x)$ leads to the momentum
constraint 
\begin{equation}
 {\cal H}_{g i}+{\cal H}_{m i}=0, \label{eqn:MomentumConstraint}
\end{equation}
where
\begin{equation}
 {\cal H}_{g i} \equiv - \frac{\delta I_g}{\delta N^i}
  = -M_{Pl}^2\sqrt{g}D^jp_{ij}, \quad
 {\cal H}_{m i} \equiv 
 -\frac{\delta I_m}{\delta N^i}. 
 \label{eqn:def-Hi}
\end{equation}
Note that the gravitational part of the momentum constraint is
determined solely by the kinetic terms and thus is totally insensitive
to the structure of higher spatial curvature terms. In particular, for
$\lambda=1$ the momentum constraint agrees with that in general
relativity.

For comparison, let us consider the case in which the matter sector
recovers spacetime diffeomorphism invariance. In this case it makes
sense to define the stress-energy tensor $T_{\mu\nu}$ of matter and then
\begin{equation}
 H_{m\perp} = \int d^3\vec{x}\sqrt{g}\ T_{\mu\nu}n^{\mu}n^{\nu}, 
  \quad 
 {\cal H}_{m i} = \frac{1}{\sqrt{g}}T_{i\mu}n^{\mu}, 
\end{equation} 
where 
\begin{equation}
 n_{\mu}dx^{\mu} = -Ndt, \quad
  n^{\mu}\partial_{\mu}= \frac{1}{N}(\partial_t-N^i\partial_i). 
  \label{eqn:unitnormal}
\end{equation}

As in general relativity, the gravitational action can be written as the
sum of kinetic terms and constraints up to boundary terms:  
\begin{equation}
 I_g = \int dt d^3\vec{x}
  \left[\pi^{ij}\partial_tg_{ij}-N^i{\cal H}_{gi}\right]
  - \int dt NH_{g\perp} + (\mbox{boundary terms}),
\end{equation}
where $\pi^{ij}$ is momentum conjugate to $g_{ij}$ given by 
\begin{equation}
 \pi^{ij} \equiv \frac{\delta I_g}{\delta (\partial_tg_{ij})}
  = M_{Pl}^2\sqrt{g}p^{ij}, \quad
  p^{ij} \equiv g^{ik}g^{jl}p_{kl}. 
\end{equation}
The Hamiltonian corresponding to the time $t$ is the sum of constraints
and boundary terms as
\begin{equation}
 H_g[\partial_t] = NH_{g\perp} + \int d^3\vec{x} N^i{\cal H}_{gi}
  + (\mbox{boundary terms}). 
\end{equation}

Finally, by variation with respect to $g_{ij}(t,x)$, we obtain the
dynamical equation 
\begin{equation}
 {\cal E}_{g ij}+{\cal E}_{m ij}=0, \label{eqn:DynamicalEquation}
\end{equation}
where
\begin{equation}
 {\cal E}_{g ij} \equiv g_{ik}g_{jl}\frac{2}{N\sqrt{g}}
  \frac{\delta I_g}{\delta g_{kl}}, \quad
 {\cal E}_{m ij} \equiv g_{ik}g_{jl}\frac{2}{N\sqrt{g}}
  \frac{\delta I_m}{\delta g_{kl}}
  = T_{ij}. 
\end{equation}
Note that the matter sector (as well as the gravity sector) should
be invariant under spatial diffeomorphism (as a part of the foliation 
preserving diffeomorphism) and thus it makes sense to define $T_{ij}$ in
general. The explicit expression for ${\cal E}_{g ij}$ is given by 
\begin{eqnarray}
 {\cal E}_{g ij} & = & M_{Pl}^2
  \left[
  -\frac{1}{N}(\partial_t-N^kD_k)p_{ij}
  + \frac{1}{N}(p_{ik}D_jN^k+p_{jk}D_iN^k)
  \right.\nonumber\\
 & & \left.
  - Kp_{ij} + 2K_i^kp_{kj}
      + \frac{1}{2}g_{ij}K^{kl}p_{kl}+ \frac{1}{2}\Lambda g_{ij} 
      - G_{ij}\right]
 + {\cal E}_{z>1 ij},
\end{eqnarray} 
where ${\cal E}_{z>1 ij}$ is the contribution from $L_{z>1}$ and
$G_{ij}$ is Einstein tensor of $g_{ij}$.

The invariance of $I_{\alpha}$ under the infinitesimal transformation 
(\ref{eqn:infinitesimal-tr}) leads to the following conservation
equations, where $\alpha$ represents $g$ or $m$. 
\begin{eqnarray}
 0 & = & N\partial_t H_{\alpha\perp}
  + \int d^3\vec{x}\left[ N^i\partial_t{\cal H}_{\alpha i}
  +\frac{1}{2}N\sqrt{g}{\cal E}_{\alpha}^{ij}\partial_tg_{ij}\right], 
 \\
 0 & = & \frac{1}{N}(\partial_t-N^jD_j)
  \left(\frac{{\cal H}_{\alpha i}}{\sqrt{g}}\right)
  + K\frac{{\cal H}_{\alpha i}}{\sqrt{g}}
  - \frac{1}{N}\frac{{\cal H}_{\alpha i}}{\sqrt{g}}D_iN^j
  - D^j{\cal E}_{\alpha ij}. 
  \label{eqn:conservation}
\end{eqnarray}

\section{Scalar graviton and the $\lambda\to 1+0$ limit}
\label{sec:scalargraviton}

\subsection{Propagating degrees of freedom}

In order to identify propagating degrees of freedom, let us consider
linear perturbations around a flat background without matter. We
can decompose the perturbation into scalar, vector and tensor parts,
according to the transformation under infinitesimal spatial
diffeomorphism. Thus, we have  
\begin{eqnarray}
 N & = & 1+A, \quad N_i = \partial_iB + n_i, \nonumber\\
  g_{ij} & = & (1+2\zeta)\delta_{ij} + 2\partial_i\partial_j h_L
  + \partial_ih_j + \partial_j h_i + h_{ij}, 
  \label{eqn:flat-plus-perturbation}
\end{eqnarray}
where $n_i$ and $h_i$ are transverse and $h_{ij}$ is transverse
traceless: 
$\delta^{ij}\partial_in_j=\delta^{ij}\partial_ih_j=0$,
$\delta^{ij}\partial_ih_{jk}=0$ and $\delta^{ij}h_{ij}=0$. 
Also, $A$ depends only on $t$ because of the projectability
condition. By fixing the gauge degrees of freedom $f(t)$ and
$\xi_i(t,\vec{x})$ as $f=-\int Adt$ and $\xi_i=-\partial_ih_L-h_i$, the 
gauge transformation (\ref{eqn:infinitesimal-tr}) leads to 
\begin{equation}
 N = 1, \quad N_i = \partial_iB + n_i, \quad 
  g_{ij} = (1+2\zeta)\delta_{ij} + h_{ij}. 
  \label{eqn:flat-plus-perturbation2}
\end{equation}

In this gauge, the momentum constraint (\ref{eqn:MomentumConstraint})
without matter is
\begin{equation}
 \partial_i
  \left[ (3\lambda-1)\dot{\zeta}
   - (\lambda-1)\partial^2B\right]
  + \frac{1}{2}\partial^2 n_i = 0,
\end{equation}
leading to
\begin{equation}
 B = \frac{3\lambda-1}{\lambda-1}\frac{\dot{\zeta}}{\partial^2},
  \quad n_i = 0, \label{eqn:sol-beta}
\end{equation}
where $\partial^2\equiv\delta^{ij}\partial_i\partial_j$ is the spatial 
Laplacian. We do not have to solve the Hamiltonian constraint since it
is an equation integrated over a whole space and thus does not reduce
the number of local physical degrees of freedom. The scalar physical
degree of freedom $\zeta$ is often called {\it scalar graviton} while
the tensor perturbation $h_{ij}$ represents the two physical degrees of
freedom of usual {\it tensor graviton}.

The time kinetic term expanded up to quadratic order is 
\begin{equation}
 I_{kin} = M_{Pl}^2
  \int dt d^3\vec{x}
  \left[\frac{3\lambda-1}{\lambda-1}\dot{\zeta}^2
   + \frac{1}{8}\delta^{ik}\delta^{jl}\dot{h}_{ij}\dot{h}_{kl}\right]
  + O(\epsilon^3) ,
  \label{eqn:kineticterm2nd}
\end{equation}
where we have introduced a small expansion parameter $\epsilon$ and
considered $\zeta$ and $h_{ij}$ as $O(\epsilon)$. In order to avoid
ghost instability, thus $\lambda$ must be either larger than $1$ or
smaller than $1/3$~\cite{Horava:2009uw}. Since general relativity has
$\lambda=1$ and we would like to recover something similar to general 
relativity in the IR, we should consider the regime
$\lambda>1$. Although the RG flow of the theory has not yet been
analyzed, a hope is that the RG flow may have a UV fixed point at
$\lambda=+\infty$ and an IR fixed point at $\lambda=1+0$ so that
$\lambda$ runs from $+\infty$ in the UV to $1+0$ in the IR.

\subsection{Dispersion relation}

Expanding the potential terms up to the second order and adding them to
(\ref{eqn:kineticterm2nd}), we obtain 
\begin{equation}
 I_g = M_{Pl}^2\int dtd^3\vec{x}
  \left[
   \frac{3\lambda-1}{\lambda-1}
   \left(\dot{\zeta}^2
    + \zeta{\cal O}_s\zeta \right)
   + \frac{1}{8}\delta^{ik}\delta^{jl}
   \left(\dot{h}_{ij}\dot{h}_{kl}+ h_{ij}{\cal O}_th_{kl}
   \right)
  \right] + O(\epsilon^3), 
\end{equation} 
where
\begin{equation}
 {\cal O}_s = \frac{\lambda-1}{3\lambda-1}
  \left(
  \frac{\Delta^3}{M_s^4}
  - \frac{\kappa_s\Delta^2}{M_s^2} - \Delta\right), \quad
 {\cal O}_t = \frac{\Delta^3}{M_t^4}
  - \frac{\kappa_t\Delta^2}{M_t^2} + \Delta.  
\end{equation} 
Here, we have introduced $M_{Pl}^2\equiv 1/(8\pi G)$, set 
$16\pi G c_8$ to unity by rescaling of the time coordinate, set
$\Lambda=0$ in order to allow the flat spacetime as a consistent
background, and defined $M_{s,t}$ and $\kappa_{s,t}$ as
\begin{eqnarray}
 M_s^{-4} & = &-2(3c_1+8c_2)M_{Pl}^{-2}, \quad
 M_t^{-4} = -2c_1M_{Pl}^{-2},  \nonumber\\
 \kappa_sM_s^{-2} & = & -2(3c_6+8c_7)M_{Pl}^{-2}, \quad
 \kappa_tM_t^{-2} = -2c_6M_{Pl}^{-2}. 
\end{eqnarray}
Thus the dispersion relation is 
\begin{equation}
 \omega^2 = \frac{\lambda-1}{3\lambda-1}
  \left(\frac{k^6}{M_s^4}
  + \frac{\kappa_s k^4}{M_s^2} - k^2\right)
 \label{eqn:scalar-dispersion}
\end{equation}
for scalar graviton, and 
\begin{equation}
 \omega^2 = \frac{k^6}{M_t^4} 
  + \frac{\kappa_t k^4}{M_t^2} + k^2
\end{equation}
for tensor graviton.

As we have already seen, the absence of ghost requires $\lambda>1$. The
dispersion relation (\ref{eqn:scalar-dispersion}) then implies that the
scalar graviton is unstable for $k$ lower than 
$\sim M$~\cite{Wang:2009yz,Blas:2009qj,Koyama:2009hc} and that the time
scale of this linear instability is
\begin{equation}
 t_L \sim \frac{1}{k}
  \sqrt{\left|\frac{3\lambda-1}{\lambda-1}\right|},
\end{equation}
As we shall see in subsection \ref{subsec:darkmatter}, the lack of local
Hamiltonian constraint leads to ``dark matter as an integration
constant'', a non-dynamical component which behaves like pressure-less
dust. As in the standard cold dark matter (CDM) scenario, the dust-like
component exhibits Jeans instability and forms large-scale structures in
the universe. The timescale of Jeans instability is 
\begin{equation}
 t_J \sim \frac{M_{Pl}}{\sqrt{\rho}}, 
\end{equation}
where $\rho$ is the energy density at the position of interest. Note
that this instability is necessary for structure formation if we
consider the dust-like component as an alternative to CDM. Thus, as far
as 
\begin{equation}
 t_L > t_J, \label{eqn:tLtJ}
\end{equation}
the linear instability of the scalar graviton does not show up. Also,
the linear instability is tamed by Hubble friction if 
\begin{equation}
 t_L > H^{-1}, \label{eqn:HtL}
\end{equation}
where $H$ is the Hubble expansion rate at the time of interest. If
either (\ref{eqn:tLtJ}) or (\ref{eqn:HtL}) is satisfied then the linear
instability of the scalar graviton does not show
up~\cite{Izumi:2009ry}. For length scales shorter than $\sim 0.01mm$, we 
do not experimentally know how gravity behaves and, thus, the linear
instability at shorter length scales would not contradict with any
experiments. Also, modes with $k$ higher than $\sim M_s$ are stable,
provided that $3c_1+8c_2<0$.

In summary, the condition under which linear instability of the scalar
graviton does not show up is 
\begin{equation}
 0 < \frac{\lambda-1}{3\lambda-1}
  < \max\left[ \frac{H^2}{k^2}, |\Phi|\right]
  \quad \mbox{for}\quad
  H < k < \min\left[M_s,\frac{1}{0.01mm}\right],
  \label{eqn:csbound}
\end{equation}
where we have introduced Newton potential $\Phi$ by 
$M_{Pl}^2k^2\Phi\sim -\rho$. Note that $\lambda$ is subject to running
under the RG flow and thus should depend on $k$, $H$ and
$\Phi$ in general. Therefore, the condition (\ref{eqn:csbound}) should
be considered as a phenomenological constraint on properties of the RG
flow.

\subsection{Breakdown of metric perturbation}

Basically, the condition (\ref{eqn:csbound}) says that $\lambda$ ($>1$)
must be sufficiently close to $1$ at low energy, while $\lambda-1$
($>0$) can be of $O(1)$ or larger at high energy. In the following we
shall show that a naive metric perturbation breaks down when $\lambda$
is close to $1$. Non-perturbatively, however, the theory is described by
a finite number of parameters, $M_{Pl}$, $\lambda$, $\Lambda$ and $c_i$
($i=1,2,\cdots,7$) if renormalizable.

A natural nonlinear extention of (\ref{eqn:flat-plus-perturbation2}) is 
\begin{equation}
 N = 1, \quad N_i = \partial_iB + n_i, \quad 
  g_{ij} = e^{2\zeta}\left[e^h\right]_{ij},
\end{equation}
where $n_i$ is transverse and $h_{ij}$ is transverse traceless:
$\delta^{ij}\partial_in_j=0$, $\delta^{ij}\partial_ih_{jk}=0$ and
$\delta^{ij}h_{ij}=0$. We shall consider $\zeta$ and $h_{ij}$ as
$O(\epsilon)$ and perform perturbative expansion with respect to
$\epsilon$.

In order to calculate the action up to cubic order, it suffices to solve 
the momentum constraint up to the first order. Thus, by substituting
(\ref{eqn:sol-beta}), we obtain 
\begin{eqnarray}
 I_{kin} & = & M_{Pl}^2
  \int dt d^3\vec{x}
  \left\{ (1+3\zeta)
   \left[\frac{3\lambda-1}{\lambda-1}\dot{\zeta}^2
		    + \frac{1}{8}\dot{h}^{ij}\dot{h}_{ij}\right]
   \right.  \nonumber\\
 & & 
  +\frac{1}{2}\zeta\partial^i(\partial_iB\partial^2B
  +3\partial^jB\partial_i\partial_jB)
  + \frac{1}{2}
  (\partial^kh_{ij}\partial_kB-3\dot{h}_{ij}\zeta)
  \partial^i\partial^jB
  \nonumber\\
 & & \left. 
      - \frac{1}{4}(\dot{h}^{ij}\partial_kh_{ij})\partial^kB
     \right\} + O(\epsilon^4),
\end{eqnarray}
where $B$ is given by (\ref{eqn:sol-beta}), and spatial indices are
raised and lowered by $\delta^{ij}$ and $\delta_{ij}$. Note that, when
written in terms of $\zeta$ and $h_{ij}$, each term in $I_{kin}$
includes exactly two time derivatives.

In order to calculate the action up to the ($n+2$)-th order
($n=1,2,\cdots$), we need to solve the momentum constraint up to $n$-th 
order. By expanding $B$ and $n_i$ as 
\begin{equation}
 B = B_1+B_2+\cdots, \quad
  n_i = n^{(1)}_i + n^{(2)}_i + \cdots, 
\end{equation}
where $B_n$ and $n^{(n)}_i$ are $O(\epsilon^n)$, and solving 
the momentum constraint perturbatively, we see that $B_n$ (and
$n^{(n)}_j$) is a sum of various terms with negative powers of 
$(\lambda-1)$ up to $(\lambda-1)^{-n}$ (and up to
$(\lambda-1)^{-(n-1)}$, respectively) and each term includes just one 
time derivative. This means that $I_{kin}$ expanded up to
$O(\epsilon^{n+2})$ includes various terms with negative powers of 
$(\lambda-1)$ up to $(\lambda-1)^{-(n+1)}$~\footnote{Terms proportional
to $(\lambda-1)^{-(n+2)}$ cancel after integration by parts.} and each
term includes exactly two time derivatives. On the other hand, terms in 
$I_{z=3,\cdots,0}$ do not include time derivatives at all and are
totally independent of $\lambda$.

Therefore, while all coefficients of potential terms for $\zeta$ and
$h_{ij}$ remain finite, many coefficients of their kinetic terms diverge 
in the limit $\lambda\to 1+0$. The divergence is worse for terms of
higher order in the perturbative expansion. This means that the naive
perturbative expansion breaks down in this limit. Here, let us stress
again that the theory is still non-perturbatively described by a finite
number of parameters, $M_{Pl}$, $\lambda$, $\Lambda$ and $c_i$ 
($i=1,2,\cdots,7$) if renormalizable.

\subsection{Non-perturbative continuity at $\lambda=1+0$}
\label{subsec:continuity}

Since the naive metric perturbation breaks down in the limit 
$\lambda\to 1+0$ , nonlinear analysis is required. In the following, for
simplicity we consider spherically symmetric, static, vacuum
configurations and show the non-perturbative continuity of the limit. In
this discussion we consider macroscopic objects and, thus, neglect
higher spatial derivative terms $I_{z=3}$ and $I_{z=2}$. Anyway,
$I_{z=3}$ and $I_{z=2}$ have well-behaved perturbative expansion and,
thus, would not spoil the continuity even if they were included. We set
the cosmological constant to zero, $I_{z=0}=0$, just for simplicity.

The lapse is required to be independent of spatial coordinates by the
projectability condition. Hence, by a space-independent time
reparametrization, we can set the lapse to unity. Then, by fixing the
gauge freedom associated with the spatial diffeomorphism, a spherically
symmetric, static configuration can be expressed as
\begin{equation}
 N = 1, \quad N_idx^i = \beta(x)dx, \quad
  g_{ij}dx^idx^j = dx^2 + r(x)^2d\Omega_2^2, 
\end{equation}
where $d\Omega_2^2$ is the line element of the unit sphere. The momentum
constraint and the $xx$-component of the dynamical equation are written
as 
\begin{eqnarray}
 \frac{\beta r''}{r}
  + (\lambda-1)
  \left[\frac{\beta'' }{2} + \frac{\beta'r'}{r} +
  \frac{\beta r''}{r} - \frac{\beta (r')^2}{r^2}\right] & = & 0, 
  \nonumber\\
 1 - (r')^2 + 2\beta\beta' rr' + 2\beta^2 rr'' + \beta^2(r')^2
  & & \nonumber\\
  \quad + (\lambda-1)
  \left[ \beta\beta'' r^2 + 2\beta^2 rr'' + 4\beta\beta' rr'
   + \frac{1}{2}(\beta')^2 r^2\right] & = & 0,
  \label{eqn:spherical-static-vacuum}
\end{eqnarray}
where a prime denotes derivative w.r.t. $x$. The
$\theta\theta$-component of the dynamical equation follows from the 
above two equations unless $r'=0$, and it is easy to show that $r'=0$ is
incompatible with the above two equations for $\lambda>1$. We shall not
impose the global Hamiltonian constraint since we are currently
interested in physics in a finite region: either staticity or spherical
symmetry is not a globally valid assumption and thus the equation
integrated over a whole space (including e.g. regions far outside the
cosmological horizon) with these assumptions at face value is not
valid. For $\beta=0$, the second equation leads to $r'=\pm 1$ and thus
allows only a trivial solution. For this reason, hereafter we assume
that $\beta\ne 0$ at least in a neighborhood of a point of interest.

It is easy to show the continuity of the $\lambda\to 1+0$ limit
explicitly. By introducing a new variable $R(x)$ by 
\begin{equation}
 R \equiv \beta^{(\lambda-1)/(2\lambda)}r,
\end{equation}
we can rewrite equations (\ref{eqn:spherical-static-vacuum}) as 
\begin{eqnarray}
 R'' + \frac{\lambda-1}{\lambda}
  \left[
   \frac{(3\lambda-1)(\beta')^2R}{4\lambda^2\beta^2}
   + \frac{(\lambda-1)\beta'R'}{\lambda\beta}
   - \frac{(R')^2}{R}\right] & = & 0, \label{eqn:R''}\\
 \frac{\beta'}{\beta} 
  - \frac{(\lambda-1)R}{4\lambda R'}\left(\frac{\beta'}{\beta}\right)^2
  + \frac{\lambda}{RR'}
   \frac{\beta^{(\lambda-1)/\lambda}
   + [(2\lambda-1)\beta^2-1](R')^2}
  {(3\lambda-1)\beta^2+(\lambda-1)}
  & = & 0. \label{eqn:beta'beta'}
\end{eqnarray}
The second equation can be solved w.r.t. $\beta'/\beta$ and there are
two branches:
\begin{eqnarray}
 \frac{\beta'}{\beta} 
  & = & \frac{1\pm\sqrt{1+4AB}}{2A},   \label{eqn:beta'}\\
  A & \equiv & \frac{(\lambda-1)R}{4\lambda R'}, \quad
  B \equiv \frac{\lambda}{RR'}
   \frac{\beta^{(\lambda-1)/\lambda}
   + [(2\lambda-1)\beta^2-1](R')^2}
  {(3\lambda-1)\beta^2+(\lambda-1)}. \nonumber
\end{eqnarray} 
The two equations (\ref{eqn:R''}) and (\ref{eqn:beta'}) provide
expressions of highest-order derivatives of $R$ and $\beta$, i.e. $R''$
and $\beta'$, as functions of ($R$, $R'$, $\beta$). For the `$-$`
branch, i.e. if we choose the `$-$` sign in (\ref{eqn:beta'}), the limit 
$\lambda\to 1+0$ of the expressions of $R$ and $\beta$ is well-defined 
as: 
\begin{equation}
 \lim_{\lambda\to 1+0}R'' = 0, \quad 
  \lim_{\lambda\to 1+0}\frac{\beta'}{\beta} 
  = \lim_{\lambda\to 1+0}
  \frac{(1-\beta^2)(R')^2-1}{2RR'\beta^2}. 
  \label{eqn:lambda1limit}
\end{equation} 
These coincide with the equations obtained by simply setting $\lambda=1$
in (\ref{eqn:R''}) and (\ref{eqn:beta'beta'}). Thus, for the `$-$`
branch, the limit $\lambda\to 1+0$ is continuous.

For comparison, let us consider general relativity with the metric ansatz 
\begin{equation}
 ds^2 = -dt^2 + \left[dx+\beta(x) dt\right]^2
  + r(x)^2d\Omega_2^2. \label{eqn:sphericalGR}
\end{equation}
Non-vanishing components of the vacuum Einstein equation $G_{\mu\nu}=0$
are 
\begin{equation}
 \beta r'' = 0, \quad 
  \beta\beta' = \frac{(1-\beta^2)(r')^2-1}{2rr'}. 
  \label{eqn:EinsteinEqSpherical}
\end{equation} 
Remember that we have assumed $\beta\ne 0$ in a neighborhood of a point
of interest. Thus, the limit $\lambda\to 1+0$ of the `$-$` branch shown
in (\ref{eqn:lambda1limit}) agrees with the Einstein equation
(\ref{eqn:EinsteinEqSpherical}). We have thus proved that, for the `$-$` 
branch, the limit $\lambda\to 1+0$ is continuous and recovers general
relativity for the metric ansatz (\ref{eqn:sphericalGR}).

\subsection{Schwarzschild solution and Newtonian limit}
\label{subsec:SchwarzschildNewtonian}

The continuity shown above, combined with Birkhoff's theorem in general
relativity, implies that the spherically symmetric, static, vacuum
solution in the '$-$' branch approaches a $3+1$ decomposition of the
Schwarzschild spacetime in the $\lambda\to 1+0$ limit. This argument
neglects higher order spatial curvature terms, $I_{z=3}$ and $I_{z=2}$,
but this is a fairly good approximation for macroscopic objects.

If we include $I_{z=3}$ and $I_{z=2}$ then in the $\lambda\to 1+0$ limit
we have 
\begin{equation}
 r' = r_1, \quad
  \frac{d}{dr}(rr_1^2\beta^2) = (r_1^2-1)
  - \sum_{z=2}^3\frac{\alpha_z(r_1)}{r^{2z}},
  \label{eqn:sphericalwithHD}
\end{equation}
where $r_1$ is a constant and $\alpha_z(r_1)$ ($z=2,3$) are constants 
depending on $r_1$ and the parameters in $I_{z=3}$ and $I_{z=2}$. Since 
the spatial metric is flat for $r_1=1$, we have 
\begin{equation}
 \alpha_z(1) = 0, \quad (z=2,3). 
  \label{eqn:alphavanish}
\end{equation}
Integrating (\ref{eqn:sphericalwithHD}), we obtain
\begin{equation}
 r_1^2\beta^2 = (r_1^2-1) + \frac{2\mu}{r}
  + \sum_{z=2}^3
  \frac{\alpha_z(r_1)}{2z-1}\frac{1}{r^{2z}}, 
\end{equation}
where $\mu$ is an integration constant~\cite{Izumi:2009ry}. For a
macroscopic object and thus for large $r$, only the first two terms are
important and, as expected, a $3+1$ decomposition of the Schwarzschild
spacetime with mass $\mu$ is recovered~\footnote{The Kerr spacetime in a
coordinate system with a unit lapse (see e.g. \cite{Doran:1999gb}) is
also a good approximate solution for a macroscopic rotating object.}: 
\begin{equation}
 r' = r_1, \quad
 r_1^2\beta^2 \simeq (r_1^2-1) + \frac{2\mu}{r}. 
 \label{eqn:lambda1Schwarzschild}
\end{equation}
The $3+1$ decomposition is characterized by the constant $r_1$. It is
noteworthy that for $r_1=1$, the solution is not just approximately but
exactly the Schwarzschild spacetime in the Painlev\'{e}-Gullstrand
coordinate system. This is because the spatial metric is flat for
$r_1=1$ and thus higher spatial curvature terms do not contribute to the
equations of motion (see (\ref{eqn:alphavanish})).

In general relativity the Newtonian limit is usually taken after going
to a gauge in which the space-dependent part of the lapse is the
Newtonian potential. How can we express the Newtonian potential in 
Ho\v{r}ava-Lifshitz gravity with the projectability condition? Actually,
all information about the Newtonian potential can be included in the
shift and the spatial metric. See the Schwarzschild solution
(\ref{eqn:lambda1Schwarzschild}) as an example. Even in general
relativity, we can choose a gauge in which the lapse is
space-independent at least locally, and in this gauge the Newtonian
potential is encoded in the shift and the spatial metric.

In Ho\v{r}ava-Lifshitz gravity, the same spacetime metric (in the sense
of general relativity) with different foliations are physically
different. Nonetheless, they are experimentally and observationally
indistinguishable from each other at low energies for the following
reason. As we all know, Lorentz invariance is a good symmetry of the
matter sector at least at low energy. It is for this reason that we need
a mechanism or symmetry to suppress Lorentz violating operators at low
energies, as already stated at the end of
subsubsection~\ref{subsubsec:scalaraction}. Therefore, although such a
mechanism or symmetry has not yet been developed and should be explored
in detail in the future, we must at the very least admit the necessity
of recovery of Lorentz symmetry in the matter sector at low energy. With
this minimal (but challenging) requirement, it is not possible to
construct low energy observables which can distinguish different
foliations of the same spacetime through motion of matter.

In summary, in Ho\v{r}ava-Lifshitz gravity with the projectability
condition, the Newtonian potential is encoded in the shift and the
spatial metric, but matter at low energy behaves as if the Newtonian
potential were expressed as the space-dependent part of the 
lapse in the ``usual'' way. Therefore, the projectability condition is not
an obstacle to expressing the Newtonian potential and taking the
Newtonian limit.

\section{Cosmological implications} 
\label{sec:cosmology}

There are a number of interesting cosmological implications of
Ho\v{r}ava-Lifshitz gravity. In this section we shall review some of
them: dark matter as an integration ``constant'' (subsection
\ref{subsec:darkmatter}), bouncing and cyclic universes (subsection
\ref{subsec:bounding}) and generation of scale-invariant cosmological
perturbation from $z=3$ scaling (subsection
\ref{subsec:scale-invariant}).

\subsection{Dark matter as an integration ``constant''}
\label{subsec:darkmatter}

\subsubsection{Structure of GR and FRW universe}

General relativity has the four-dimensional diffeomorphism invariance as
its fundamental symmetry. As a result, there are four local constraints:
one Hamiltonian constraint and three momentum constraints at each
spatial point and at each time. The constraints are preserved by
dynamical equations. Thus, we can solve dynamical equations without
worrying about constraints, provided that constraints are satisfied at
initial time.

Now, let us consider the flat FRW spacetime
\begin{equation}
 ds^2 = -dt^2 + a^2(t)d\vec{x}^2, \label{eqn:flatFRW}
\end{equation}
as a simple example. It is supposed that this metric approximates
overall behavior of our patch of the universe inside the Hubble
horizon. The Hamiltonian constraint leads to the Friedmann equation
\begin{equation}
 3H^2 = 8\pi G \rho, 
\end{equation}
where $G$ is Newton's constant, $H\equiv\dot{a}/a$ is the Hubble
expansion rate and $\rho$ is the total energy density of matter contents
of the universe. Equations of motion of matter lead to the conservation 
equation
\begin{equation}
 \dot{\rho} + 3H(\rho+P) = 0, \label{eqn:FRWconservation}
\end{equation}
where $P$ is the total pressure of the matter contents. The momentum
constraint is trivial because of the symmetry of the FRW spacetime. The
dynamical equation 
\begin{equation}
  -(2\dot{H}+3H^2)
  = 8\pi G P \label{eqn:FRWdynamical}
\end{equation}
follows from the Friedmann equation and the conservation equation, and
thus we do not consider it as an independent equation.

\subsubsection{Structure of HL gravity and FRW universe}
\label{subsubsec:HLFRW}

The fundamental symmetry of Ho\v{r}ava-Lifshitz gravity is the
invariance under the foliation preserving diffeomorphism
(\ref{eqn:foliation-preserving-diffeo}), which is $3$-dimensional
spatial diffeomorphism plus space-independent time
reparametrization. Consequently, contrary to general relativity, the
theory has $3$ local constraints and $1$ global constraint: $3$ 
momentum constraints at each spatial point at each time and $1$ 
Hamiltonian constraint integrated over a whole space at each time. Of 
course, the constraints are preserved by dynamical equations. Thus, we
can solve dynamical equations without worrying about constraints,
provided that constraints are satisfied at initial time.

Now let us consider the flat FRW spacetime (\ref{eqn:flatFRW}), or 
\begin{equation}
 N=1, \quad N^i=0, \quad g_{ij} = a(t)^2\delta_{ij}, 
  \label{eqn:HLflatFRW}
\end{equation}
in Ho\v{r}ava-Lifshitz gravity. Again, the FRW spacetime is supposed to 
approximate overall behavior of our patch of the universe inside the
Hubble horizon. This means that the global Hamiltonian constraint, which
is an integral over a whole space including regions far outside the
Hubble horizon, does not apply to our system within the horizon. Thus, 
the lack of local Hamiltonian constraint implies that there is no
Friedmann equation and that we should consider the dynamical equation 
\begin{equation}
-\frac{3\lambda-1}{2}(2\dot{H}+3H^2) = 8\pi G P
 \label{eqn:FRWdynamicaleq}
\end{equation}
as an independent equation. Here, note that higher spatial curvature
terms do not contribute to the equation because of the spatial
flatness. Equations of motion for matter leads to the conservation
equation (\ref{eqn:FRWconservation}) at least at low energy, provided
that the local Lorentz invariance is restored in the matter sector at
low energy as required by many experimental and observational data (see
discussion in the second-to-the-last paragraph of subsection
\ref{subsec:SchwarzschildNewtonian}). At high energy, 
however, the matter sector does not have to satisfy the conservation
equation and thus the equation of motion for matter generally leads to 
\begin{equation}
 \dot{\rho} + 3H(\rho+P) = -Q, \label{eqn:non-conservation}
\end{equation}
where $Q$ represents the amount of energy non-conservation. Note that
$Q\to 0$ at low energy. The two equations (\ref{eqn:FRWdynamicaleq}) and
(\ref{eqn:non-conservation}) are sufficient to describe the evolution of
our system. Indeed, it is easy to obtain the first integral of the
dynamical equation: 
\begin{equation}
 \frac{3(3\lambda-1)}{2}H^2
  = 8\pi G \left[\rho+\frac{C(t)}{a^3}\right], 
  \label{eqn:FrEQwDM}
\end{equation}
where
\begin{equation}
 C(t) \equiv C_0 + \int_{t_0}^t Q(t') a^3(t')dt',
  \label{eqn:defC(t)}
\end{equation}
and $C_0=C(t_0)$ is an integration constant. Since $Q\to 0$ at low
energy, 
\begin{equation}
 C(t) \to const. \quad \mbox{at low energy}. 
\end{equation}
The first integral (\ref{eqn:FrEQwDM}) looks like Friedmann equation
but, intriguingly, the extra term ($\propto C(t)/a^3$) behaves like dark
matter at low energy. This term is not real matter but gravitationally 
behaves like pressure-less dust. Thus, in Ho\v{r}ava-Lifshitz gravity,
something like dark matter emerges as an integration ``constant'' at
least in flat FRW background at low energy. Note that
(\ref{eqn:defC(t)}) describes how the ``dark matter'' is generated in
the early universe: even with $C_0=0$ and $t_0=-\infty$, we have
non-vanishing $C(t)$ at late time.

\subsubsection{General case in the IR}

We now show that the dark matter as an integration ``constant'' emerges
at low energy in more general situation.

Low energy behavior of the theory is described by the IR action
(\ref{eqn:IRaction}). This looks like the Einstein Hilbert action with
the ADM decomposition if $\lambda=1$. Hence, we set $\lambda=1$ in the
discussion below, hoping that in the near future we can show that
$\lambda=1$ is a stable IR fixed point of the RG flow.

The Hamiltonian constraint is then of the form
\begin{equation}
 \int d^3\vec{x}\sqrt{g}
  \left[ G^{(4)}_{\mu\nu} + \Lambda g^{(4)}_{\mu\nu}
   - 8\pi G T_{\mu\nu} \right]n^{\mu}n^{\nu} = 0,
\end{equation}
where $g^{(4)}_{\mu\nu}$ is the four-dimensional spacetime metric
defined in (\ref{eqn:ADMmetric}), $G^{(4)}_{\mu\nu}$ is the Einstein 
tensor of $g^{(4)}_{\mu\nu}$, $n^{\mu}$ is the unit vector normal to the
constant time hypersurface defined in (\ref{eqn:unitnormal}), and
$T_{\mu\nu}$ is the energy momentum tensor of matter. This is an
equation integrated over a whole space including regions far outside the
cosmological horizon, and thus does not restrict physics inside our
patch of the universe. On the other hand, the momentum constraint 
\begin{equation}
  \left[ G^{(4)}_{i\mu} + \Lambda g^{(4)}_{i\mu}
   - 8\pi G T_{i\mu} \right]n^{\mu} = 0, 
  \label{eqn:momentum-constraintIR}
\end{equation}
and dynamical equations 
\begin{equation}
 G^{(4)}_{ij} + \Lambda g^{(4)}_{ij} - 8\pi G T_{ij} = 0,
  \label{eqn:dynamical-eqIR}
\end{equation}
are local equations.

Interestingly, it is possible to give a general solution to these local
equations. For this purpose let us define $T^{dark}_{\mu\nu}$ by 
\begin{equation}
 T^{dark}_{\mu\nu} \equiv 
  (8\pi G)^{-1}
  \left[ G^{(4)}_{\mu\nu}
   +\Lambda g^{(4)}_{\mu\nu} - 8\pi G T_{\mu\nu}
       \right].  \label{eqn:defTHL}
\end{equation} 
The momentum constraint (\ref{eqn:momentum-constraintIR}) and the
dynamical equations (\ref{eqn:dynamical-eqIR}) are rewritten as
\begin{equation}
 T^{dark}_{i\mu} n^{\mu} = 0, \quad T^{dark}_{ij} = 0,
\end{equation} 
meaning that the time-space components and the space-space components of
$T^{dark}_{\mu\nu}$ vanish. Only the time-time component remains and
thus $T^{dark}_{\mu\nu}$ should be proportional to $n_{\mu}n_{\nu}$: 
\begin{equation}
 T^{dark}_{\mu\nu} = \rho_{dark}n_{\mu}n_{\nu}, 
\end{equation} 
where the scalar $\rho_{dark}$ can in general depend on both time and
spatial coordinates. This is exactly of the form of pressure-less
dust and thus behaves like dark matter. It is easy to show that the
vector $n^{\mu}$ defined in (\ref{eqn:unitnormal}) follows the geodesic
equation 
\begin{equation}
 n^{\nu}\nabla_{\nu}n^{\mu}=0. 
  \label{eqn:geodesiceq}
\end{equation}
Also, by taking the divergence of the definition (\ref{eqn:defTHL}) of
$T^{dark}_{\mu\nu}$, we can show that $\rho_{dark}$ satisfies the
conservation equation
\begin{equation}
 n^{\mu}\partial_{\mu}\rho_{dark} + K\rho_{dark} = 0,
  \label{eqn:rhoHLconservation}
\end{equation}
provided that the real matter sector recovers the local Lorentz
invariance in the IR and thus satisfies the energy conservation 
$n^{\mu}\nabla^{\nu}T_{\mu\nu}=0$ at low energy. In more general cases
the right hand side of (\ref{eqn:rhoHLconservation}) obtains
non-vanishing contributions from higher spatial curvature terms,
deviation of $\lambda$ from unity, and energy non-conservation of
matter.

As a consistency check, let us apply the conservation equation
(\ref{eqn:rhoHLconservation}) to the flat FRW spacetime
(\ref{eqn:HLflatFRW}). In this case, (\ref{eqn:rhoHLconservation}) is
reduced to $\partial_t\rho_{dark}+3H\rho_{dark}=0$ and thus
$\rho_{dark}\propto a^{-3}$. This reproduces the scale factor dependence
of the last term in (\ref{eqn:FrEQwDM}) with $C(t)=const$.

In summary, we have shown that gravity equations of motion in
Ho\v{r}ava-Lifshitz gravity at low energy with $\lambda=1$ is written as
\begin{equation}
 G^{(4)}_{\mu\nu} + \Lambda g^{(4)}_{\mu\nu}
   = 8\pi G \left[T_{\mu\nu}+\rho_{dark}n_{\mu}n_{\nu}\right]. 
   \label{eqn:modifiedEinsteineq}
\end{equation}
This {\it modified Einstein equation} includes a built-in component
which behaves like dark matter, as an inevitable consequence of the 
projectability condition. The ``dark matter velocity vector'' $n^{\mu}$
follows the geodesic equation (\ref{eqn:geodesiceq}) and the ``dark
matter energy density'' $\rho_{dark}$ satisfies the conservation
equation (\ref{eqn:rhoHLconservation}). In the Newtonian limit the
modified Einstein equation (\ref{eqn:modifiedEinsteineq}) reduces to the
Poisson equation with the built-in ``dark matter'' included. Note that,
as already discussed in subsection \ref{subsec:SchwarzschildNewtonian},
the Newtonian potential is encoded not in the lapse but in the shift and
the spatial metric.

\subsection{Bouncing and cyclic universes}
\label{subsec:bounding}

Higher curvature terms in the action are expected to play important
roles in the early universe. In this section we consider the FRW
universe with spatial curvature
\begin{equation}
 N=1, \quad N^i=0, \quad 
  g_{ij}dx^idx^j = a(t)^2
  \left[\frac{dr^2}{1-Kr^2} + r^2d\Omega_2^2\right], 
\end{equation}
and see that higher curvature terms drastically change the evolution of
the early universe. In particular, bouncing universes and cyclic
universes are allowed as regular solutions in Ho\v{r}ava-Lifshitz
gravity.

\subsubsection{Modified Friedmann equation with higher curvature terms}

As already stated in subsubsection \ref{subsubsec:HLFRW}, the FRW
spacetime is just an approximation to describe overall behavior of our
patch of the universe inside the Hubble horizon, and thus the global
Hamiltonian constraint does not restrict the dynamics of our approximate
FRW universe inside the horizon. (This is evident in the presence of 
superhorizon fluctuations.) Instead, we should consider the dynamical
equation as an independent equation, 
\begin{equation}
-\frac{3\lambda-1}{2}(2\dot{H}+3H^2) = 
 8\pi G P
 - \frac{\alpha_3K^3}{a^6}
 - \frac{\alpha_2K^2}{a^4}
 + \frac{K}{a^2} - \Lambda, 
\end{equation}
where
\begin{equation}
\alpha_3 = 192\pi G (c_3+3c_4+9c_5), \quad
 \alpha_2 = 32\pi G (c_6+3c_7). 
\end{equation}

By using the definition (\ref{eqn:non-conservation}) of energy
non-conservation $Q$, we can easily obtain the first integral of the
dynamical equation, 
\begin{equation}
\frac{3(3\lambda-1)}{2}H^2 = 
 8\pi G \left[\rho + \frac{C(t)}{a^3} \right]
 - \frac{\alpha_3K^3}{a^6}
 - \frac{3\alpha_2K^2}{a^4}
 - \frac{3K}{a^2} + \Lambda , \label{eqn:FrEQwDMwK}
\end{equation}
where $C(t)$ is defined in (\ref{eqn:defC(t)}). This is a
straightforward generalization of (\ref{eqn:FrEQwDM}) and includes
contributions from the spatial curvature $K$. As in (\ref{eqn:FrEQwDM}),
the term proportional to $C(t)/a^3$ behaves like dark matter at low
energy as $C(t)\to const$. In the early universe, i.e. for small $a$,
the curvature cubic term ($\propto K^3/a^6$) plays important roles.

In order to see qualitative behavior of the system, let us rewrite the
first integral (\ref{eqn:FrEQwDMwK}) in the form of the energy
conservation equation for a non-relativistic particle moving in a 
$1$-dimensional potential as 
\begin{equation}
 \frac{1}{2}\dot{a}^2 + \frac{2}{3\lambda-1}V(a) = 0,
\end{equation}
where
\begin{equation}
 V(a) = 
  \frac{\alpha_3K^3}{6a^4}+\frac{\alpha_2K^2}{2a^2}
   + \frac{K}{2} - \frac{\Lambda}{6} a^2
   - \frac{4\pi G}{3}\left[\rho a^2+\frac{C(t)}{a}\right]. 
   \label{eqn:defV(a)}
\end{equation}
The shape of the potential $V(a)$ completely determines the behavior of
the system.

\subsubsection{Simple examples}

\begin{figure}[htbp]
 \begin{minipage}{.42\linewidth}
  \begin{center}
\includegraphics[width=\linewidth]{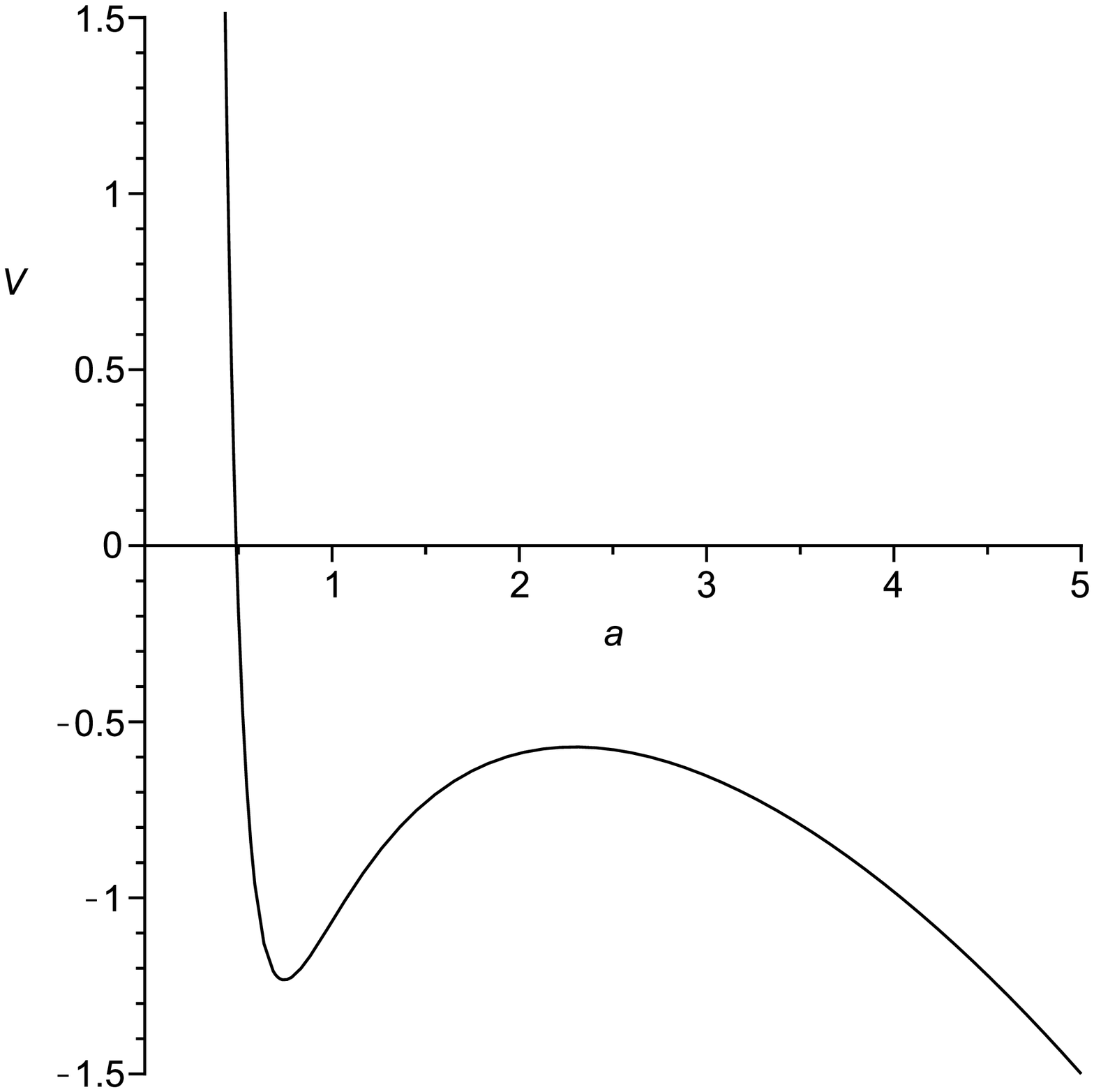}
  \end{center}
  \caption{$V(a)$ for $\Lambda=0.4$ and $8\pi GC=10$. 
  If the universe is initially contracting then it bounces and expands.}
  \label{fig:bouncing}
 \end{minipage} 
 \begin{minipage}{0.05\linewidth} 
  \begin{center}
  \end{center}
 \end{minipage} 
 \begin{minipage}{.42\linewidth}
  \begin{center}
\includegraphics[width=\linewidth]{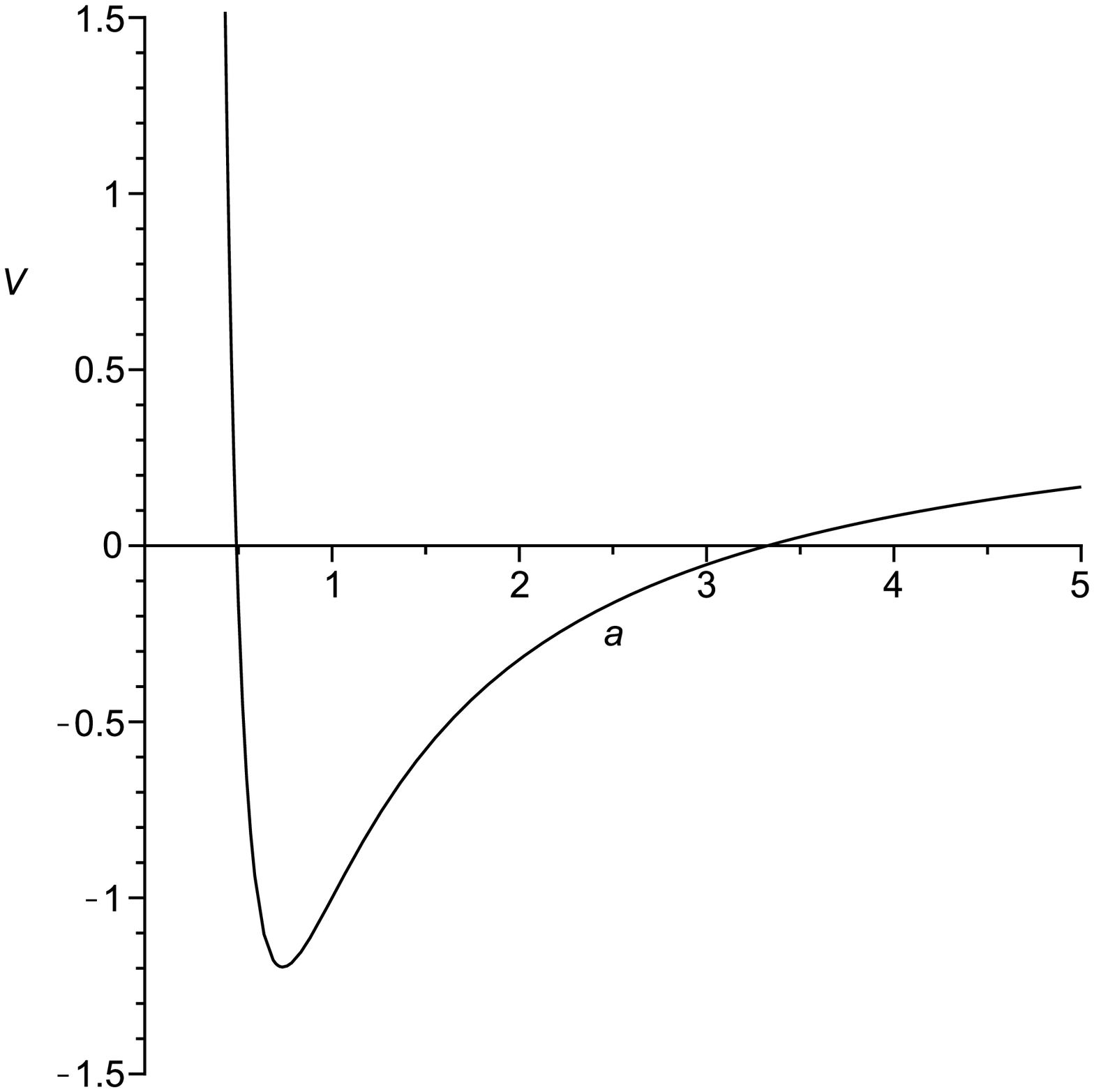}
  \end{center}
  \caption{$V(a)$ for $\Lambda=0$ and $8\pi GC=10$. 
  This allows a series of expansion and contraction, i.e. a cyclic
  universe.}
  \label{fig:cyclic}
 \end{minipage} 
\end{figure}

\begin{figure}[htbp]
 \begin{minipage}{.42\linewidth}
  \begin{center}
\includegraphics[width=\linewidth]{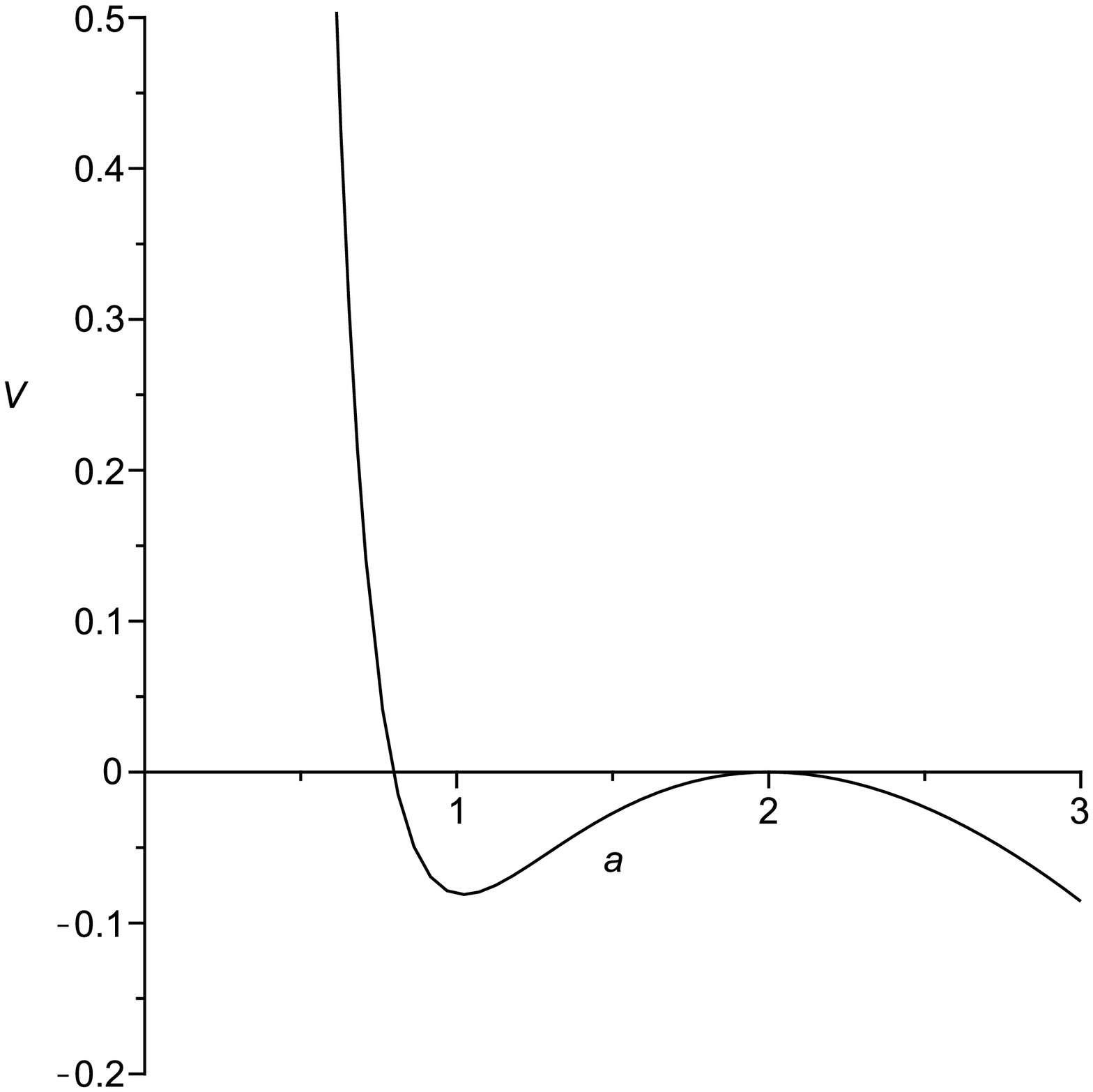}
  \end{center}
  \caption{$V(a)$ for $\Lambda=15/64$ and $8\pi GC=17/4$. 
  This allows a static universe at $a=2$ but it is unstable. A bouncing
  universe is also allowed.}
  \label{fig:unstable}
 \end{minipage} 
 \begin{minipage}{0.05\linewidth} 
  \begin{center}
  \end{center}
 \end{minipage} 
 \begin{minipage}{.42\linewidth}
  \begin{center}
\includegraphics[width=\linewidth]{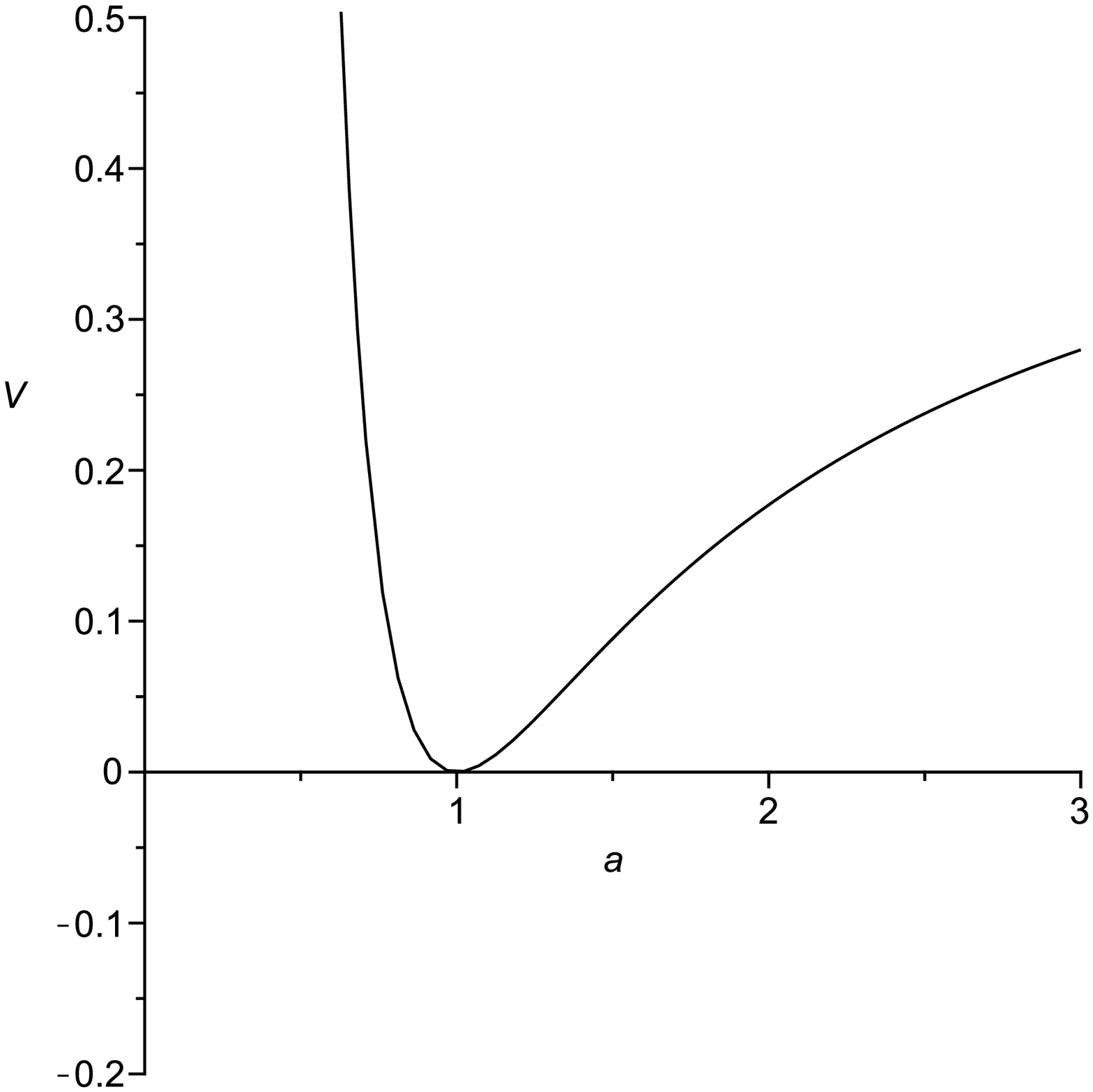}
  \end{center}
  \caption{$V(a)$ for $\Lambda=0$ and $C=4$. 
  This allows a stable static universe at $a=1$.}
  \label{fig:stable}
 \end{minipage} 
\end{figure}

Let us now consider some simple examples. For simplicity we set
$\alpha_3=1$, $\alpha_2=0$, $K=1$, $\rho=0$, $C=const$. We still have
freedom to choose values of $\Lambda$ and $C$. We show four examples of
the $1$-dimensional potential (\ref{eqn:defV(a)}): a bouncing universe
(Figure \ref{fig:bouncing}), a cyclic universe (Figure
\ref{fig:cyclic}), an unstable static universe (Figure
\ref{fig:unstable}) and a stable static universe (Figure
\ref{fig:stable}). See \cite{Maeda:2010ke} for more examples with
$\rho=0$ and $C=const$.


\subsection{Scale-invariant cosmological perturbations from $z=3$ scaling}
\label{subsec:scale-invariant}

One of the essential ingredients of Ho\v{r}ava-Lifshitz gravity is the
anisotropic scaling with the dynamical critical exponent 
$z\ge 3$. Indeed, it is this property that makes the theory
power-counting renormalizable and attractive as a candidate for the
theory of quantum gravity. There are interesting cosmological
implications of the anisotropic scaling. In this section we show that
the anisotropic scaling with the minimal $z$, i.e. $z=3$, leads to a new
mechanism for the generation of scale-invariant cosmological
perturbations. Intriguingly, this mechanism works even without
inflation.

\subsubsection{Usual story with $z=1$}

Before explaining the new mechanism, let us remind ourselves of the
usual story with $z=1$. 

Cosmological perturbations are analyzed by perturbative expansion
around a FRW background. In the linearized level, perturbations are
Fourier expanded and the evolution of each mode is characterized by the 
frequency $\omega$ defined by the dispersion relation
\begin{equation}
 \omega^2 = c_s^2\frac{k_c^2}{a^2},
  \label{eqn:dispersionz=1}
\end{equation}
where $c_s$ is the sound speed, $k_c$ is the comoving wave number and
$a$ is the scale factor of the universe. For simplicity we assume that
the time dependence of $c_s$, if any, is slow compared with the
cosmological time scale $H^{-1}$, where $H=\dot{a}/a$ is the Hubble
expansion rate. (For example, $c_s$ is identically $1$ for a canonical
scalar field with any potential.)

If a mode of interest satisfies $\omega^2\gg H^2$ then the evolution of 
the mode is not affected by the expansion of the universe and the mode
just oscillates. When $\omega^2\ll H^2$, on the other hand, the
expansion of the universe is so rapid that the Hubble friction freezes
the mode and the mode stays almost constant. Generation of cosmological
perturbations from quantum fluctuations is nothing but the oscillation
followed by the freeze-out. Therefore, the condition for generation of
cosmological perturbations is 
\begin{equation}
 \frac{d}{dt}\left(\frac{H^2}{\omega^2}\right) > 0. 
  \label{eqn:osc-freeze}
\end{equation}
With the $z=1$ dispersion relation (\ref{eqn:dispersionz=1}), this
condition is equivalent to $\ddot{a}>0$ for expanding universe
($\dot{a}>0$). Therefore, if $z=1$ then generation of cosmological
perturbations from quantum fluctuations requires accelerated expansion
of the universe, i.e. inflation. For example, for power law expansion
$a\propto t^p$, $p>1$ is required.

Observational data of the cosmic microwave background strongly indicates
that the primordial cosmological perturbations have an almost
scale-invariant spectrum. It is easy to see that the scale-invariance
also requires inflation. From the scaling (\ref{eqn:scaling}) with
$s=1$, the amplitude of quantum fluctuations of the scalar field should
be proportional to the energy scale of the system. In cosmology the
energy scale is set by the Hubble expansion rate $H$. Thus, we expect
that 
\begin{equation}
 \delta \phi \propto H. \label{eqn:deltaphiz=1}
\end{equation}
Since cosmological perturbations with different scales are generated at
different times, the scale-invariance is nothing but the constancy of
the right hand side of (\ref{eqn:deltaphiz=1}). Noting that
$H=\dot{a}/a$, this implies the exponential expansion of the universe 
$a\propto\exp(Ht)$, namely inflation.

We have seen that, for $z=1$, both the generation of cosmological
perturbations and the scale-invariance of generated perturbations
require the existence of an inflationary epoch in the early universe.

\subsubsection{The story in the UV with $z=3$}

The condition (\ref{eqn:osc-freeze}) for generation of cosmological
perturbations is valid irrespective of the dispersion relation. In
Ho\v{r}ava-Lifshitz gravity, to realize the anisotropic scaling
(\ref{eqn:anisotropic-scaling}), the dispersion relation for
a physical degree of freedom in the UV should be 
\begin{equation}
 \omega^2 = M^2\times\left(\frac{k_c^2}{M^2a^2}\right)^z,
  \label{eqn:dispersion}
\end{equation}
where $M$ is some energy scale. By substituting this to the condition
(\ref{eqn:osc-freeze}) we obtain $d^2(a^z)/dt^2>0$ for expanding
universe ($\dot{a}>0$). Since $z\geq 3$ in Ho\v{r}ava-Lifshitz gravity
at high energy, generation of cosmological perturbations from quantum
fluctuations does not require accelerated expansion of the universe,
i.e. inflation. For example, power law expansion $a\propto t^p$ with
$p>1/z$ satisfies the condition.

In this way, the anisotropic scaling provides a solution to the horizon
problem. Essential reason for this is that perturbations freeze-out not
at the Hubble horizon but at the {\it sound horizon}, defined by 
$\omega\sim H$. The physical radius of sound horizon is thus 
$\sim (M^{z-1}H)^{-1/z}$. In the UV epoch ($H\gg M$), the sound horizon 
is far outside the Hubble horizon and can therefore accommodate scales
much longer than the Hubble horizon size. In order to stretch
microscopic scales to cosmological scales, we just need to have a long
enough expansion history (satisfying the condition $d^2(a^z)/dt^2>0$) in
the UV epoch. Note that $M$ is not a cutoff scale of a low energy
effective theory but is just the scale at which the theory starts
exhibiting the anisotropic scaling, provided that Ho\v{r}ava-Lifshitz
gravity is UV complete.

For general $z$, the formula (\ref{eqn:scaling-dim}) implies that the
amplitude of quantum fluctuations of $\phi$ should be 
\begin{equation}
 \delta\phi \sim M\times \left(\frac{H}{M}\right)^{\frac{3-z}{2z}},
  \label{eqn:deltaphi}
\end{equation}
where $M$ is defined through the dispersion relation
(\ref{eqn:dispersion}). This is of course consistent with the well-known
result (\ref{eqn:deltaphiz=1}) for $z=1$ and the result in ghost 
inflation 
$\delta\phi\sim (M^3H)^{1/4}$~\cite{ArkaniHamed:2003uy,ArkaniHamed:2003uz} 
for $z=2$. On the other hand, in Ho\v{r}ava-Lifshitz gravity with the
minimal value of $z$ ,i.e. $z=3$, (\ref{eqn:deltaphi}) is reduced to
$\delta\phi\sim M$, implying that the amplitude of quantum fluctuations
does not depend on the Hubble expansion rate. This means that the 
spectrum of cosmological perturbations in Ho\v{r}ava-Lifshitz gravity
with $z=3$ is automatically scale-invariant even without inflation.

\subsubsection{A simple model}

We have shown that the anisotropic scaling with $z=3$ naturally leads to
a new mechanism for generation of scale-invariant cosmological
perturbations. As a simple implementation of the mechanism, let us
consider a free scalar field described by the action 
\begin{equation}
 I = \frac{1}{2}\int dt d^3\vec{x}N\sqrt{g}
  \left[ \frac{1}{N^2}(\partial_t\phi-N^i\partial_i\phi)^2
   + \phi{\cal O}\phi \right],
\end{equation}
where
\begin{equation}
 {\cal O} = \frac{\Delta^3}{M^4}
  - \frac{\kappa\Delta^2}{M^2}
  + c_{\phi}^2\Delta - m_{\phi}^2, \quad 
  \Delta = g^{ij}D_iD_j
\end{equation}
This is a covariantized version of (\ref{eqn:free-scalar-flat}).

In the UV, the first term in ${\cal O}$ is dominant and the scalar field
action exhibits the $z=3$ scaling. In this regime it is easy to find the 
mode function in a flat FRW background as~\cite{Mukohyama:2009gg}
\begin{equation}
 \phi_{\vec{k}_c} = \frac{e^{i\vec{k}_c\cdot\vec{x}}}{(2\pi)^3}
  \times 2^{-1/2}k_c^{-3/2}M
  \exp\left(-i\frac{k_c^3}{M^2}\int\frac{dt}{a^3}\right),
\end{equation}
where $a$ is the scale factor, $t$ is the proper time, $\vec{k}_c$ is
the comoving wave number and $k_c=|\vec{k}_c|$. Note that this is not
just WKB approximation but actually exact and applicable to both
subhorizon and superhorizon scales in any background $a(t)$, provided
that the first term in ${\cal O}$ is dominant. The mode function
approaches a constant value in the $a\to\infty$ limit if and only if the
integral $\int^{\infty}dt/a^3$ converges. For power-law expansion
$a\propto t^p$, this condition is satisfied if $p>1/3$, and agrees with
the condition for the freeze-out after oscillation discussed after
(\ref{eqn:dispersion}). Provided that the integral converges, the
power-spectrum is calculated as 
\begin{equation}
 {\cal P}_{\phi} = 
  \frac{k_c^3}{2\pi^2}
  \left| (2\pi)^3\phi_{\vec{k}_c}\right|^2
  = \left(\frac{M}{2\pi}\right)^2. 
\end{equation}
This is manifestly scale-invariant in accord with the general argument
after (\ref{eqn:deltaphi}). In this way, scale-invariant cosmological
perturbations of the scalar field can be generated even without
inflation.

After scales of interest exit the sound horizon, cosmological
perturbations of the scalar field can be converted to curvature
perturbations by either curvaton mechanism or modulated decay of heavy
particles or/and oscillating fields. For example, it is possible to
suppose that the scalar field $\phi$ itself plays the role of a
curvaton~\cite{Mukohyama:2009gg}. When the Hubble expansion rate
becomes as low as $m_{\phi}$, $\phi$ starts rolling and eventually
decays to radiation. Perturbations of $\phi$ are converted to those of
radiation energy density and thus curvature perturbations.

In the IR, the first two terms in ${\cal O}$ can be neglected and the
usual $z=1$ scaling is recovered. In this epoch, unless the universe is
in an inflationary phase, physical scales re-enter the horizon as
usual.

\section{Summary and discussions}
\label{sec:summary}

We have reviewed basic construction and cosmological implications of a
power-counting renormalizable theory of gravitation recently proposed by
Ho\v{r}ava. While there are many fundamental issues to be addressed in
the future, it is interesting to investigate cosmological
implications.

Since the high energy behavior of Ho\v{r}ava-Lifshitz gravity is very
different from general relativity, there is a possibility that the
theory does not exactly recover general relativity at low energy. As
reviewed in subsection \ref{subsec:darkmatter}, this is indeed the case
and the theory can instead mimic general relativity plus dark
matter. The constraint algebra in this theory is smaller than general
relativity since the time slicing is synchronized with the ``dark
matter rest frame'' in the theory level. In subsection
\ref{subsec:bounding} we have shown that higher spatial curvature terms
in the action drastically change the evolution of the early universe. We
have derived modified Friedmann equation with higher spatial curvature
terms and have shown some simple examples, including bouncing and cyclic
universes. The anisotropic scaling at high energy is one of essential
ingredients of the theory since the power-counting renormalizability
stems from it. In subsection \ref{subsec:scale-invariant} we have
reviewed a new mechanism for generation of cosmological perturbations
based on the anisotropic scaling. This mechanism can solve the horizon
problem and generate scale-invariant cosmological perturbations even
without inflation.

In Sec.~\ref{sec:scalargraviton} we have commented on some issues
related to the scalar graviton and the $\lambda\to 1+0$ limit, where
$\lambda$ is a parameter in the kinetic action. We have explicitly seen
that the naive metric perturbation breaks down for the scalar graviton
in the $\lambda\to 1+0$ limit. However, this does not necessarily imply
the loss of predictability. Actually, for spherically-symmetric, static, 
vacuum configurations we have proved that the limit is
non-perturbatively continuous and safely recovers general relativity.

Now let us compile a list of some important open questions. 
\begin{itemize}
\item 
      Renormalizability must be shown beyond power-counting
      argument. (See \cite{Orlando:2009en} for discussion about
      renormalizability of the theory with the detailed balance
      condition.) 
\item 
      The RG flow of the theory must be analyzed. In particular, it is
      very important to see whether $\lambda=1$ is an IR fixed point or
      not. If it is the case then we would like to know whether the RG
      flow can satisfy the condition (\ref{eqn:csbound}) or not. 
\item 
      We have to develop mechanisms or symmetries to suppress Lorentz
      violating operators in the matter sector at low energies. Perhaps,
      embedding into a larger theory is needed. One such possibility is
      related to supersymmetry~\cite{GrootNibbelink:2004za}. 
\item 
      Is there an analogue of Vainshtein
      effect~\cite{Vainshtein:1972sx}? In subsection
      \ref{subsec:continuity}, non-perturbative continuity of the 
      $\lambda\to 1+0$ limit was shown only for spherically-symmetric, 
      static, vacuum configurations. We need to consider more general
      situations in order to see how general the non-perturbative
      continuity is. 
\item
     In \cite{Mukohyama:2009tp}, based on exact results in some simple
     cases, it was conjectured that there is no caustic for constant
     time hypersurfaces. We need to provide evidences for this
     conjecture in more general situations if a proof is
     difficult. Perhaps, numerical simulations similar to those in
     \cite{ArkaniHamed:2005gu} are necessary. 
\item
     In \cite{Izumi:2009ry} it was proved that a spherically-symmetric
     solution should include a time-dependent region near the center. 
     On the other hand, as shown in Sec.~\ref{sec:scalargraviton} of the
     present article, the vacuum region far from the center recovers the
     standard Schwarzschild geometry. Since the size of the dynamical
     region is expected to be of the fundamental scale, the dynamical
     nature of the central region is not really relevant for macroscopic
     objects such as astrophysical stars. Microscopically, however, this
     could be rather significant . We would like to know, e.g. the
     typical size of the dynamical region and the motion of its
     boundary. 
\item
     As already stressed in \cite{Mukohyama:2009tp}, we need to know if
     microscopic lumps of ``dark matter as an integration constant''
     can play the role of particles in usual dark matter models from 
     macroscopic viewpoint. Interactions among them such as collisions
     and bounces need to be understood. At astrophysical scales, we need
     to see if collective behavior of a group of large number of
     microscopic lumps can more or less mimic behavior of a cluster of 
     particles with velocity dispersion and vorticity. Clearly, detailed
     investigation is necessary to understand rich dynamics of ``dark
     matter'' from microscopic to macroscopic scales. 
\item
     As shown in (\ref{eqn:defC(t)}), ``dark matter as an integration
     constant'' is generated in the early universe even if it vanishes
     initially. This formula can be applied to superhorizon
     perturbations. Given a concrete model of the matter sector,
     therefore, it is straightforward to estimate the typical amplitude
     and spectrum of the ``dark matter''. If a single physical degree of
     freedom is responsible for both the source term $Q$ in
     (\ref{eqn:defC(t)}) and generation of curvature perturbations then
     it should be possible to realize adiabatic initial conditions for
     the late time evolution of perturbations. (See
     \cite{Kobayashi:2009hh} for classical late time evolution.) It is
     worthwhile investigating this possibility in details. 
\item
     The mechanism reviewed in subsection \ref{subsec:scale-invariant}
     generates scale-invariant cosmological perturbations without a need 
     for inflation. It would be interesting to see whether
     renormalization effects such as anomalous dimension can break the
     exact scale-invariance and explain the observed spectral tilt. 
\item
     In the early universe, it is expected that $\lambda$ should deviate
     from $1$ under the RG flow and that the scalar graviton can be
     treated perturbatively. Since the scalar graviton has the $z=3$
     anisotropic scaling in the UV, it should also obtain the
     scale-invariant cosmological
     perturbations~\cite{Chen:2009jr}. Provided that $\lambda=1$ is a
     stable IR fixed point of the RG flow, as the universe expands and
     the Hubble expansion rate decreases, $\lambda$ approaches $1$ and
     the perturbative treatment of the scalar graviton becomes
     invalid~\footnote{Thus, the conventional cosmological perturbation 
     scheme~\cite{Gong:2010xp} probably breaks down for the scalar
     graviton in the $\lambda\to 1+0$ limit. Again, this does not
     necessarily imply loss of predictability but requires nonlinear
     analysis.}. However, the result in subsection
     \ref{subsec:continuity} suggests that the $\lambda\to 1+0$ limit 
     may be non-perturbatively continuous. A natural question is then
     ``how to convert the scale-invariant cosmological perturbations of 
     the scalar graviton to observables such as cosmic microwave
     background anisotropies and matter power spectrum?'' 
\end{itemize}

Since Ho\v{r}ava's original proposal in January 2009, several extensions
appeared in the literature. Blas, et.al.~\cite{Blas:2009qj} proposed an
extension without the projectability condition by including spatial
derivatives of the lapse in the action. More recent proposal by 
Ho\v{r}ava and Melby-Thompson~\cite{Horava:2010zj} respects the
projectability condition but the fundamental symmetry of the theory is
larger than the original one.

Throughout this article, we have considered the minimal theory, i.e. the
original theory with the projectability condition but without extension 
of the symmetry. Whether this minimal theory is viable is still an open
question and crucially depends on non-perturbative nature of the scalar
graviton (see subsection \ref{subsec:continuity}) and properties of the
RG flow (see the condition (\ref{eqn:csbound}) ).

\section*{Acknowledgments}

The author would like to thank Keisuke Izumi, Takeshi Kobayashi, Satoshi
Maeda, Kazunori Nakayama, Tetsuya Shiromizu, Fuminobu Takahashi and
Shuichiro Yokoyama for fruitful collaboration on this subject. He is
grateful to Frans Klinkhamer, Massimo Porrati, Valery Rubakov, Misao
Sasaki, Masaru Shibata and Takahiro Tanaka for useful discussions. 
The work of the author is supported by JSPS Grant-in-Aid for Young
Scientists (B) No.~17740134, JSPS Grant-in-Aid for Creative Scientific
Research No.~19GS0219, MEXT Grant-in-Aid for Scientific Research on
Innovative Areas No.~21111006, JSPS Grant-in-Aid for Scientific Research
(C) No.~21540278, and the Mitsubishi Foundation. This work was supported
by World Premier International Research Center Initiative.


\end{document}